# Joint probability density with radial, tangential, and perturbative forces

Jae-Won Jung[1], Sung Kyu Seo[1], Sungchul Kwon[2], Kyungsik Kim[1,3,*]

[1]*DigiQuay Company Ltd., Seocho-gu Seoul 06552, Republic of Korea*
[2]*Department of Physics, Catholic University of Korea, Bucheon 14662, Korea*
[3]*Department of Physics, Pukyong National University, Busan 608-737, Korea*

We study the Fokker-Planck equation for an active particle with both the radial and tangential forces and the perturbative force. We find the solution of the joint probability density. In the limit of $t \gg \tau$ and for $\tau = 0$ domains, the mean squared radial velocity for an active particle leads to a super-diffusive distribution, while the mean squared tangential velocity with both the radial and tangential forces and the perturbative force behaviors as the Gaussian diffusion. Compared with the self-propelled particle, the mean squared tangential velocity is matched with the same value to the time $\sim t^2$, while the mean squared radial velocity is the same as the time $\sim t$ in $t \gg \tau$ and for $\tau = 0$. The moment $\mu_{2,2}$ for the joint probability density with the radial and tangential forces scales as the time $\sim t^3$ in the limit of $t \gg \tau$ and for $\tau = 0$, while that for the joint probability density with the perturbative force is proportional to the time $\sim t^4$ in $t \ll \tau$.

## 1. Introduction

The motion of Brownian particle is an object of interest for many times, as is well known in the stochastic process. Orenstein and Uhlenbeck [1] have discussed three Brownian motions with harmonic force treated separately, and the formula of such motions is recognized to take the same form given by Smoluchowski [2]. The Mori-Zwanzig research [3] has yielded a generalized Langevin equation with viscous force and colored noise on the influence of complex and mixed environments. Baxion and Zwanzig [4] have applied the techniques of non-equilibrium statistical mechanics to a simple nonlinear Duffing oscillator with linear and cubic terms in the restoring force. A Brownian particle larger than the small surrounding particles moves with the fluctuating force caused by collisions of surrounding particles and viscous force interrupting and decelerating its motion. In this way, a Brownian particle, as described by the fluctuation-dissipation theorem [5, 6], is called the passive particle with no destination, while the active particle that goes towards its desired destination. The derivation of Fokker- Planck equation and the exact solution of joint probability density with an exponentially correlated Gaussian force have been studied and analyzed by Heinrichs [7].

The problems of solving the force equation, generalized Langevin equation, with harmonic, perturbative, and other forces, have been studied and analyzed by the scientific researchers in diverse fields. In addition, such problems in several sorts of mediums include the statistical values of various diffusion constants, first passage times, correlation coefficients, memory functions, and so on. Chaudhuri and Dhar [8] recently have studied an active Brownian particle within a harmonic trap in the presence of the translational diffusion, and they presented an explicit calculation of the several moments at arbitrary times and their evolution to the steady state. In addition, Patel and Chaudhuri [9] have shown that the steady-state velocity distribution exhibits a re-entrant crossover from the passive Gaussian to the active non-Gaussian behaviors and constructed a corresponding the phase diagram using the exact expression of the kurtosis. Over five decades, despite the deluge of the various papers that have been written on the calculation and the analysis of the Brownian motion described the statistical properties of classical and modern systems, there have been satisfactorily the theories, the computer-simulations, and the experimental devices with the utility, the generality, and the precision from the colloid, the run-and-tumble particle, the micro-swimmer to the flocking models, the schools of fish, the swarms of insects, the slime molds, the herds of wildebeest [10-12]. Features between passive and active particles have been described and analyzed by the different fluctuations and statistical properties [13,14]. Examples of the active systems variously include the

Brownian motors [15], the motile cells [16-18], the macroscopic animals [19,20], the artificial self-propelled particles [21,22], and the interacting collective particles [23-25]. Examples of the driving force in the active matter research are also the automated digital tracking [26-28] and the active granular and colloidal systems [29-33]. We will approximately solve the differential equation of the joint probability density made from classical circular motion. In our model, the equation of motion with the directional and tangential forces is altered, and may be similarly expressed as an equation of the self-propelled motion. In Section 2, after deriving the technique of the Fokker-Planck equation, we use the double Fourier transform to solve the joint distribution function with both the directional and the tangential forces and the perturbative force. The approximate solution of the probability density is found in three domains, i.e. $t \ll \tau$, $t \gg \tau$, and $\tau = 0$, where $\tau$ denotes the correlation time. The kurtosis, the correlation coefficient, and the moment from the moment equation are calculated and analyzed. In Section 3, we finally provide and account for a conclusion summarizing the key findings.

## 2. Joint probability density with the radial and tangential forces

### 2-1. Fokker-Planck equation

We firstly consider an active particle that have an accelerated circular motion in a plane polar coordinate. The differential displacement and the velocity are, respectively, $d\vec{s} = dr\hat{r} + rd\theta\hat{\theta}$ and $d\vec{s}/dt = \dot{r}\hat{r} + r\dot{\theta}\hat{\theta}$, where $\hat{r}$ and $\hat{\theta}$ are, respectively, the unit vector of radial direction and of tangential direction, respectively. The motion of force for the accelerated circular motion acting on an active particle is simply written as $m\frac{d\vec{v}}{dt} = m\frac{d\vec{v}_r}{dt} + m\frac{d\vec{v}_\theta}{dt} = -mr\omega^2\hat{r} + mr\alpha\hat{\theta}$, i.e. the sum of the centripetal force and the tangential force. We introduce that the positive force by energy flowing from the tangential direction denotes $ma_2 v_\theta$, and that the viscous force denotes $-mr_1 v_r$ in the radial direction. The positive force also denotes $ma_1 v_r$ by energy flowing from the radial direction and the accelerated force denoted $mr_2 \dot{v}_\theta$ in the tangential direction. An exponentially correlated Gaussian force $g(t)$ is given as

$$<g(t)> = 0, \quad <g(t)g(t')> = rk_B T \frac{1}{\tau}\exp[-\frac{|t-t'|}{\tau}] = \frac{g_0^2}{2\tau}\exp[-\frac{|t-t'|}{\tau}]. \tag{1}$$

Here, $r$ denotes the viscous coefficient, $k_B$ the Boltzmann constant, $T$ the temperature, $g_0$ the coupling strength, and $\tau$ the characteristic time.

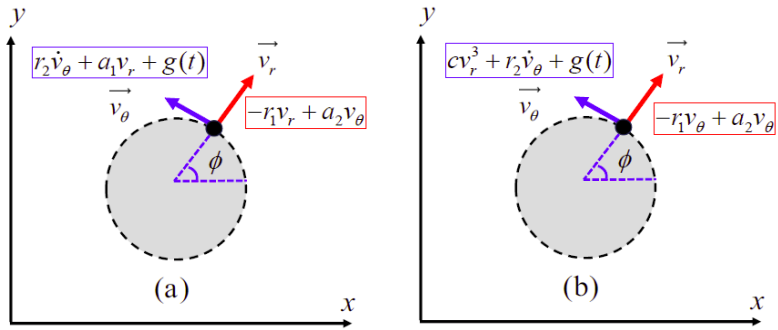

**Figure 1.** Plots of an active particle with (a) the radial and tangential forces and (b) perturbative force. The forces in the two vector directions are $-r_1 v_r + a_2 v_\theta$, $-r_1 v_\theta + a_2 v_\theta$ in the radial direction and $r_2 \dot{v}_\theta + a_1 v_r + g(t)$, $cv_r^3 + r_2 \dot{v}_\theta + g(t)$ in tangential direction, respectively. Here, $\vec{v}_r$, $\vec{v}_\theta$ denote the unit vectors of the radial, tangential directions.

Thus, Figure 1(a) is shown for the sum of such forces we introduced. The equation of motion for an active particle with the radial and tangential forces is expressed in terms of

$$\frac{d}{dt}v_r = -r_1 v_r + a_2 v_\theta, \tag{2}$$

$$\frac{d}{dt}v_\theta = r_2 \dot{v}_\theta + a_1 v_r + g(t). \tag{3}$$

Here, the dimensionless mass is $m=1$. The forces are reduced as $-mr\omega^2 = -mr_1 v_r$ and $mr\alpha = mr_2 \dot{v}_\theta$, respectively.

Secondly, we introduce a differential equation for an active particle with the perturbative force in the two following equations. That is,

$$\frac{d}{dt}v_r = -r_1 v_\theta + a_2 v_\theta = (-r_1 + a_2)v_\theta, \tag{4}$$

$$\frac{d}{dt}v_\theta = cv_r^3 + r_2 \dot{v}_\theta + g(t). \tag{5}$$

In Eq. (4), the first term of the first equality, the viscous force, denotes $-r_1 v_\theta$. The first term of right-handed side of Eq. (5), the perturbative force, denotes $cv_r^3(t)$ in Figure 2(b).

From now on, we drive the Fokker-Planck equation. It is assumed that the joint probability density of the particle be initially at rest at time $t$ is defined by $<\delta(v_r - v_r(t))\delta(v_\theta - v_\theta(t))>$. By taking time derivatives of the joint probability density, we have

$$\frac{\partial}{\partial t}P(v_r(t),v_\theta(t),t) = -\frac{\partial}{\partial v_r}<\frac{\partial v_r}{\partial t}\delta(v_r - v_r(t))\delta(v_\theta - v_\theta(t))>$$
$$-\frac{\partial}{\partial v_\theta}<\frac{\partial v_\theta}{\partial t}\delta(v_r - v_r(t))\delta(v_\theta - v_\theta(t))>. \tag{6}$$

From Eq. (1) and Eq. (2), the joint probability density $P(v_r(t),v_\theta(t),t) \equiv P(v_r,v_\theta,t)$ with the correlated Gaussian force is derived as

$$\frac{\partial}{\partial t}P(v_r,v_\theta,t) = [r_1\frac{\partial}{\partial v_r}v_r - a_2 v_\theta \frac{\partial}{\partial v_r} - \frac{\alpha}{1-r_2}b(t)\frac{\partial^2}{\partial v_r \partial v_\theta} - \frac{a_1}{1-r_2}v_r\frac{\partial}{\partial v_\theta} + \frac{\alpha}{1-r_2}a(t)\frac{\partial^2}{\partial v_\theta^2}]P(v_r,v_\theta,t). \tag{7}$$

Here $\alpha = g_0^2/2$, $a(t) = [1-\exp(-t/\tau)]$, and $b(t) = (t+\tau)\exp(-t/\tau) - \tau$.

## 2-2. $P(v_r,t)$ and $P(v_\theta,t)$ with the radial and tangential forces in the short-time domain

In order to compute the joint probability density, it is expedient to use the double Fourier transform of the distribution for $v_r$ and $v_\theta$ is given by

$$P(\xi,t) = \int_{-\infty}^{+\infty} dv_r \exp(-i\nu v_r)P(v_r,t), \tag{8}$$

$$P(\nu,t) = \int_{-\infty}^{+\infty} dv_\theta \exp(-i\nu v_\theta)P(v_\theta,t). \tag{9}$$

The double Fourier transform Eq. (8) and Eq. (9) govern the probability density of Eq. (7) for $v_r$, $v_\theta$. Then, we can derive $P(\xi,\nu,t)$ as

$$\frac{\partial}{\partial t}P(\xi,\nu,t) = [-r_1\xi\frac{\partial}{\partial \xi} + a_2\xi\frac{\partial}{\partial \nu} + \frac{a_1}{1-r_2}\nu\frac{\partial}{\partial \xi} + \frac{\alpha}{1-r_2}[b(t)\nu\xi - a(t)\nu^2]]P(\xi,\nu,t). \tag{10}$$

By separating the variables $v_r$ and $v_\theta$ in Eq. (10), the Fourier transform of the Fokker-Planck equation is reduced as

$$\frac{\partial}{\partial t}P(\xi,t) = (\frac{a_1}{1-r_2}\nu - r_1\xi)\frac{\partial}{\partial \xi}P(\xi,t) + \frac{1}{2}\frac{\alpha}{1-r_2}[b(t)\nu\xi - a(t)\nu^2]P(\xi,t) + bP(\xi,t) \tag{11}$$

$$\frac{\partial}{\partial t}P(\nu,t) = a_2\xi\frac{\partial}{\partial \nu}P(\nu,t) + \frac{1}{2}\frac{\alpha}{1-r_2}[b(t)\nu\xi - a(t)\nu^2]P(\nu,t) - bP(\nu,t). \tag{12}$$

Here, $b$ is the separation constant. As we take $\frac{\partial}{\partial t}P(\xi,t) = 0$ in the steady state, we get $P^{st}(\xi,t)$ from Eq. (11) as

$$P^{st}(\xi,t) = \exp[\frac{\alpha}{2a_1\nu}[a(t)\nu^2\xi - \frac{b(t)}{2}\nu\xi^2] + \frac{\alpha r_1(1-r_2)}{2(a_1\nu)^2}[\frac{a(t)}{2}\nu^2\xi^2 - \frac{b(t)}{3}\nu\xi^3] - [\frac{b(1-r_2)}{a_1\nu}\xi + \frac{b(1-r_2)^2}{(a_1\nu)^2}\frac{\xi^3}{3}]]. \tag{13}$$

We assume that $[\frac{a_1\nu}{(1-r_2)} - r_1\xi]^{-1} \cong \frac{(1-r_2)}{a_1\nu}[1 + \frac{r_1(1-r_2)\xi}{a_1\nu}]$ in Eq. (13). In order to find the solutions of joint functions for $\xi$ from $P_\xi(\xi,t) \equiv Q(\xi,t)P^{st}(\xi,t)$, we obtain the distribution functions in $t \ll \tau$. That is,

$$P(\xi,t) = Q(\xi,t)P^{st}(\xi,t)$$
$$= Q(\xi,t)\exp[\frac{\alpha}{2a_1\nu}[a(t)\nu^2\xi - \frac{b(t)}{2}\nu\xi^2] + \frac{\alpha r_1(1-r_2)}{2(a_1\nu)^2}[\frac{a(t)}{2}\nu^2\xi^2 - \frac{b(t)}{3}\nu\xi^3]$$
$$-[\frac{b(1-r_2)}{a_1\nu}\xi + \frac{b(1-r_2)^2}{(a_1\nu)^2}\frac{\xi^3}{3}]], \tag{14}$$

$$Q(\xi,t) = R(\xi,t)Q^{st}(\xi,t)$$
$$= R(\xi,t)\exp[\frac{\alpha(1-r_2)}{2(a_1\nu)^2}[\frac{a'(t)}{2}\nu^2\xi^2 - \frac{b'(t)}{6}\nu\xi^3] + \frac{\alpha r_1(1-r_2)^2}{2(a_1\nu)^3}[\frac{a'(t)}{6}\nu^2\xi^3 - \frac{b'(t)}{12}\nu\xi^4]], \tag{15}$$

$$R(\xi,t) = S(\xi,t)R^{st}(\xi,t)$$
$$= S(\xi,t)\exp[\frac{\alpha(1-r_2)^2}{2(a_1v)^3}[\frac{a''(t)}{6}v^2\xi^3 - \frac{b''(t)}{24}v\xi^4] + \frac{\alpha r_1(1-r_2)^3}{2(a_1v)^4}[\frac{a''(t)}{24}v^2\xi^3 - \frac{b''(t)}{60}v\xi^5]], \quad (16)$$

$$S(\xi,t) = T(\xi,t)S^{st}(\xi,t) = T(\xi,t)\exp[-\frac{\alpha(1-r_2)^3}{2(a_1v)^4}\frac{b'''(t)}{120}v\xi^5 - \frac{\alpha r_1(1-r_2)^4}{2(a_1v)^5}\frac{b'''(t)}{360}v\xi^6]. \quad (17)$$

Neglecting terms proportional to $1/\tau^3$, we put $T(\xi,t)$ as $T(\xi,t) = \Theta[t + \xi/[a_1(1-r_2)^{-1}v - r_1\xi]]$, which we choose $T(\xi,t)$ as an arbitrary function of variable $t + \xi/[a_1(1-r_2)^{-1}v - r_1\xi]$. We consequently find

$$P(\xi,t) = T(\xi,t)S^{st}(\xi,t)R^{st}(\xi,t)Q^{st}(\xi,t)P^{st}(\xi,t)$$
$$= \Theta[t + [\xi/[a_1(1-r_2)^{-1}v - r_1\xi]]S^{st}(\xi,t)R^{st}(\xi,t)Q^{st}(\xi,t)P^{st}(\xi,t)$$
$$= \exp[-\frac{\alpha r_1^2 t^3}{a_1}[1 - \frac{1}{8}\frac{t}{\tau}]\xi^2 - \frac{\alpha r_1^2 t^4}{2(1-r_2)}[1 + \frac{1}{r_1 t}]v\xi - \frac{5\alpha a_1 t^4}{48(1-r_2)^2\tau}[1 - \frac{12}{5}\frac{\tau}{t}]v^2]. \quad (18)$$

By using the inverse Fourier transform, $P(v_r,t)$ is presented by

$$P(v_r,t) = [4\pi\frac{\alpha r_1^2 t^3}{a_1}[1 - \frac{1}{8}\frac{t}{\tau}]]^{-1/2}\exp[-\frac{a_1 v_r^2}{4\alpha r_1^2 t^3}[1 - \frac{1}{8}\frac{t}{\tau}]^{-1}]. \quad (19)$$

The mean squared velocity for $P(v_r,t)$ is given by

$$<v_r^2> = \frac{2\alpha r_1^2 t^3}{a_1}[1 - \frac{1}{8}\frac{t}{\tau}]. \quad (20)$$

For the short-time domain, we use Eq. (12) in the tangential direction. In steady state, introducing $\frac{\partial}{\partial t}P(v,t) = 0$ for $v$ and $P(v,t) \to P^{st}(v,t)$, then we calculate the joint probability density as

$$P^{st}(v,t) = \exp[\frac{\alpha}{2(1-r_2)a_2\xi}[\frac{a(t)}{3}v^3 - \frac{b(t)}{2}\xi v^2] + \frac{b}{a_2\xi}v]. \quad (21)$$

In order to find the solution of joint function for $v$ from $P(v,t) \equiv Q(v,t)P^{st}(v,t)$, we obtain the probability density in the short-time domain $t \ll \tau$. That is,

$$P(v,t) = Q(v,t)P^{st}(v,t) = Q(v,t)\exp[\frac{\alpha}{2(1-r_2)a_2\xi}[\frac{a(t)}{3}v^3 - \frac{b(t)}{2}\xi v^2] + \frac{b}{a_2\xi}v], \quad (22)$$

$$Q(v,t) = R(v,t)Q^{st}(v,t) = R(v,t)\exp[\frac{\alpha}{2(1-r_2)(a_2\xi)^2}[\frac{a'(t)}{12}v^4 - \frac{b'(t)}{6}\xi v^3]], \quad (23)$$

$$R(v,t) = S(v,t)R^{st}(v,t) = R(v,t)\exp[\frac{\alpha}{2(1-r_2)(a_2\xi)^3}[\frac{a''(t)}{60}v^5 - \frac{b''(t)}{24}\xi v^4]], \quad (24)$$

$$S(v,t) = T(v,t)S^{st}(v,t) = T(v,t)\exp[-\frac{\alpha}{2(1-r_2)(a_2\xi)^4}\frac{b'''(t)}{120}\xi v^5]. \quad (25)$$

By neglecting terms proportional to $1/\tau^3$, we take the joint probability density as an arbitrary function of a variable $t + (v/a_2\xi)$. That is, the arbitrary function $T(v,t)$ is set to $T(v,t) = \Theta[t + (v/a_2\xi)]$. By expanding their derivatives to second order in powers of $t/\tau$, we obtain the expression for $P(v,t)$ after some cancellations. We consequently derive and calculate that

$$P(v,t) = T(v,t)S^{st}(v,t)R^{st}(v,t)Q^{st}(v,t)P^{st}(v,t) = \Theta[t + (v/a_2\xi)]R(v,t)Q(v,t)P^{st}(v,t)$$
$$= \exp[-\frac{5\alpha a_2 t^4}{48(1-r_2)\tau}[1 - \frac{12}{5}\frac{\tau}{t}]\xi^2 - \frac{\alpha t^3}{12(1-r_2)\tau}[1 - 6\frac{\tau}{t}]\xi v - \frac{3\alpha t^2}{8(1-r_2)a_2\tau}[1 - \frac{2}{9}\frac{t}{\tau}]v^2] \quad (26)$$

with

$$\Theta(u) = \exp[[-\frac{\alpha a_2}{120(1-r_2)\tau^2}\xi^2 u^5]_{S^{st}(v,t)} + [[\frac{\alpha a_2^2}{120(1-r_2)\tau^2} - \frac{\alpha a_2}{48(1-r_2)\tau}]\xi^2 u^5]_{R^{st}(v,t)}$$
$$-[[\frac{\alpha a_2}{24(1-r_2)\tau} + \frac{\alpha a_2 t}{12(1-r_2)\tau}]\xi^2 u^4]_{q^{st}(v,t)} - [[\frac{\alpha a_2^2 t}{6(1-r_2)\tau} - \frac{\alpha a_2 t}{4(1-r_2)}]\xi^2 u^3]_{P^{st}(v,t)}]. \quad (27)$$

Here, $\exp[f(z)]_{S^{st}(v,t)}$ is the calculated value of $S^{st}(v,t)$ in Eq. (25). For the inverse Fourier transform, $P(v_\theta,t)$ is presented by

$$P(v_\theta,t) = [\pi\frac{3\alpha t^2}{2(1-r_2)a_2\tau}[1 - \frac{2}{9}\frac{t}{\tau}]]^{-1/2}\exp[-\frac{(1-r_2)a_2\tau v_\theta^2}{3\alpha t^2}[1 - \frac{2}{9}\frac{t}{\tau}]^{-1}]. \quad (28)$$

The mean squared velocity for $P(v_\theta,t)$ is given by

$$<v_\theta^2> = \frac{3\alpha t^2}{4(1-r_2)a_2\tau}[1 - \frac{2}{9}\frac{t}{\tau}]. \quad (29)$$

## 2-3. $P(v_r,t)$ and $P(v_\theta,t)$ with the radial and tangential forces in the long-time domain

In the long-time domain $t \gg \tau$, $P_\xi(\xi,t)$ from Eq. (11) is given by

$$\frac{\partial}{\partial t} P_\xi(\xi,t) \cong \frac{1}{2}\frac{\alpha}{(1-r_2)}[b(t)v\xi - a(t)v^2]P_\xi(\xi,t) \cdot \qquad (30)$$

The above equation is easily calculated as

$$P_\xi(\xi,t) = \exp[\frac{\alpha}{2(1-r_2)}\int[b(t)v\xi - a(t)v^2]dt] \cdot \qquad (31)$$

Here $\int a((t)dt = t - \tau$ and $\int b((t)dt = -\tau t$. In steady state, we also calculate from $P_\xi(\xi,t) = Q(\xi,t)P^{st}(\xi,t)$ like

$$Q^{st}(\xi,t) = \exp[-\frac{\alpha}{2(1-r_2)}\int[b(t)v\xi - a(t)v^2]dt] \cdot \qquad (32)$$

In the steady state, we use the same result of Eq. (13) as $P(\xi,t) \to P^{st}(\xi,t)$. As an arbitrary function $R(\xi,t)$ becomes $R(\xi,t) = \Theta[t + \xi/[\alpha_1 v - r_1\xi]]$, we derive and calculate $P(\xi,t)$ as

$$P(\xi,t) = R(\xi,t)Q^{st}(\xi,t)P^{st}(\xi,t) = \Theta[t + \xi/[a_1(1-r_2)^{-1}v - r_1\xi]]Q^{st}(\xi,t)P^{st}(\xi,t)$$

$$= \exp[-\frac{\alpha r_1^3(1-r_2)t^2}{4a_1}[1-\frac{2}{r_1 t}]\xi^2 - \frac{\alpha r_1\tau t^2}{2(1-r_2)}[1+\frac{1}{r_1 t}]v\xi - \frac{\alpha r_1 t^2}{4(1-r_2)}[1+\frac{2}{r_1 t}]v^2] \cdot \qquad (33)$$

By using $Q^{st}(\xi,t)$ of Eq. (32) and $P^{st}(v,t)$ of Eq. (21) for $v$, we similarly get

$$P(v,t) = R(v,t)Q^{st}(v,t)P^{st}(v,t) = \Theta[t + v/a_2\xi]Q^{st}(v,t)P^{st}(v,t)$$

$$= \exp[-\frac{\alpha a_2^2(1-r_2)t^3}{6(1-r_2)}[1+\frac{3}{2}\frac{\tau}{a_2 t}]\xi^2 - \frac{\alpha a_2 t^2}{2(1-r_2)}[1+\frac{\tau}{a_2 t}]v\xi - \frac{\alpha t}{2(1-r_2)}v^2] \cdot \qquad (34)$$

From Eq. (33) and Eq. (34), $P(v_r,t)$ and $P(v_\theta,t)$ using the inverse Fourier transform are, respectively, presented by

$$P(v_r,t) = [\pi \frac{\alpha r_1^3(1-r_2)t^2}{a_1}[1-\frac{2}{r_1 t}]]^{-1/2} \exp[-\frac{a_1 v_r^2}{\alpha r_1^3(1-r_2)t^2}[1-\frac{2}{r_1 t}]^{-1}], \qquad (35)$$

$$P(v_\theta,t) = [2\pi\frac{\alpha t}{(1-r_2)}]^{-1/2} \exp[-\frac{(1-r_2)v_\theta^2}{2\alpha t}] \cdot \qquad (36)$$

After a little calculation, the mean squared velocities for $P(v_r,t)$ and $P(v_\theta,t)$ are, respectively, given by

$$<v_r^2> = \frac{\alpha r_1^3(1-r_2)t^2}{2a_1}[1-\frac{2}{r_1 t}], \qquad (37)$$

$$<v_\theta^2> = \frac{\alpha t}{(1-r_2)} \cdot \qquad (38)$$

## 2-4. $P(v_r,t)$ and $P(v_\theta,t)$ with the radial and tangential forces for $\tau = 0$

From now on, we will obtain the solution of joint probability density for $\tau = 0$. The Fourier transforms of Fokker-Planck equation in Eq. (11) and Eq. (12) are, respectively, reduced to

$$\frac{\partial}{\partial t} P(\xi,t) = [\frac{a_1}{1-r_2}v - r_1\xi]\frac{\partial}{\partial \xi} P(\xi,t) - \frac{1}{2}\frac{\alpha}{1-r_2}v^2 P(\xi,t), \qquad (39)$$

$$\frac{\partial}{\partial t} P(v,t) = a_2\xi\frac{\partial}{\partial v} P(v,t) - \frac{1}{2}\frac{\alpha}{1-r_2}v^2 P(v,t) \cdot \qquad (40)$$

Here, $a(t) = 1$ and $b(t) = 0$ for time domain $\tau = 0$. The exact solutions for $\xi$ and $v$ are, respectively, derived as

$$P(\xi,t) = \Theta[t + \xi/[a_1(1-r_2)^{-1}v - r_1\xi]]P^{st}(\xi,t)$$

$$= \Theta[t + \xi/[a_1(1-r_2)^{-1}v - r_1\xi]]\exp[\frac{\alpha}{2a_1}[\frac{a_1}{(1-r_2)}tv^2 + \frac{1}{2}r_1 tv]], \qquad (41)$$

$$P(v,t) = \Theta[t + \frac{1}{\alpha_2\xi}]P^{st}(v,t) = \Theta[t + v/a_2\xi]\exp[\frac{\alpha}{6(1-r_2)a_2\xi}v^3] \cdot \qquad (42)$$

After a little calculation, Eq. (41) and (42) are, respectively, obtained by

$$P(\xi,t) = \exp[-\frac{\alpha r_1^3 t^2}{4a_1^2(1-r_2)}[1-\frac{2}{r_1 t}]\xi^2 - \frac{\alpha r_1 t}{2a_1(1-r_2)}[1-\frac{1}{r_1 t}]v\xi - \frac{\alpha r_1 t^2}{4(1-r_2)}[1+\frac{2}{r_1 t}]v^2], \qquad (43)$$

$$P(v,t) = \exp[-\frac{\alpha a_2^2 t^3}{6(1-r_2)}\xi^2 - \frac{\alpha a_2 t^2}{2(1-r_2)}v\xi - \frac{\alpha t}{2(1-r_2)}v^2] \cdot \qquad (44)$$

By using the inverse Fourier transform, $P(v_r,t)$ and $P(v_\theta,t)$ are, respectively, presented by

$$P(v_r,t) = [\pi \frac{\alpha r_1^3 t^2}{a_1^2(1-r_2)}[1-\frac{2}{r_1 t}]]^{-1/2} \exp[-\frac{a_1^2(1-r_2)v_r^2}{\alpha r_1^3 t^2}[1-\frac{2}{r_1 t}]^{-1}], \tag{45}$$

$$P(v_\theta,t) = [2\pi \frac{\alpha t}{(1-r_2)}]^{-1/2} \exp[-\frac{(1-r_2)v_\theta^2}{2\alpha t}]. \tag{46}$$

The mean squared velocities for $P(v_r,t)$ and $P(v_\theta,t)$ are, respectively, calculated by

$$<v_r^2> = \frac{\alpha r_1^3 t^2}{2a_1^2(1-r_2)}[1-\frac{2}{r_1 t}], \tag{47}$$

$$<v_\theta^2> = \frac{\alpha t}{(1-r_2)}. \tag{48}$$

## 3. Joint probability density with the perturbative force

### 3-1. $P(v_r,t)$ and $P(v_\theta,t)$ with the perturbative force in the short-time domain

In this subsection, we will consider the equation of motion of Eq. (2-4) and Eq. (2-5) for an active particle with perturbative force. The Fourier transforms of the variable-separated Fokker-Planck equation are, respectively, represented in terms of

$$\frac{\partial}{\partial t}P(\xi,t) = -\frac{c}{(1-r_2)}vD_\xi^2\frac{\partial}{\partial \xi}P(\xi,t) + \frac{\alpha}{2(1-r_2)}[b(t)\xi v - a(t)v^2]P(\xi,t) + eP(\xi,t), \tag{49}$$

$$\frac{\partial}{\partial t}P(v,t) = [(a_2-r_1)\xi]\frac{\partial}{\partial v}P(v,t) + \frac{\alpha}{2(1-r_2)}[b(t)\xi v - a(t)v^2]P(v,t) - eP(v,t), \tag{50}$$

where $D_\xi = d/d\xi$ and $e$ denotes the separation constant. The Fourier transform of the probability density for $\xi$ is given by $P(\xi,t) = Q(\xi,t)P^{st}(\xi,t)$, so we continuously obtain the distribution functions from Eq. (49) in the short-time domain. That is,

$$P(\xi,t) = Q(\xi,t)P^{st}(\xi,t) = Q(\xi,t)\exp[\frac{\alpha}{2cvD_\xi^2}[\frac{b(t)}{2}v\xi^2 - a(t)v^2\xi] + \frac{e(1-r_2)}{cvD_\xi^2}\xi], \tag{51}$$

$$Q(\xi,t) = R(\xi,t)Q^{st}(\xi,t) = R(\xi,t)\exp[-\frac{\alpha(1-r_2)}{2(cvD_\xi^2)^2}[\frac{b'(t)}{6}v\xi^3 - \frac{a'(t)}{2}v^2\xi^2]], \tag{52}$$

$$R(\xi,t) = S(\xi,t)R^{st}(\xi,t) = S(\xi,t)\exp[\frac{\alpha(1-r_2)^2}{2(cvD_\xi^2)^3}[\frac{b''(t)}{24}v\xi^4 - \frac{a''(t)}{6}v^2\xi^3]], \tag{53}$$

$$S(\xi,t) = T(\xi,t)S^{st}(\xi,t) = T(\xi,t)\exp[-\frac{\alpha(1-r_2)^3}{2(cvD_\xi^2)^4}\frac{b'''(t)}{120}v\xi^5]. \tag{54}$$

Hence, we derive $P(\xi,t)$ after a little calculation as

$$P(\xi,t) = T(\xi,t)S^{st}(\xi,t)R^{st}(\xi,t)Q^{st}(\xi,t)P^{st}(\xi,t)$$
$$= \Theta[t+\xi/[c(1-r_2)^{-1}vD_\xi^2]]S^{st}(\xi,t)R^{st}(\xi,t)Q^{st}(\xi,t)P^{st}(\xi,t). \tag{55}$$

We use Eq. (50) for the tangential direction in the limit of $t \ll \tau$. In steady state, introducing $\frac{\partial}{\partial t}P(v,t)=0$ for $v$ and calculating $P^{st}(v,t)$, we continuously find

$$P(v,t) = Q(v,t)P^{st}(v,t) = Q(v,t)\exp[\frac{\alpha}{2(1-r_2)(a_2-r_1)\xi}[\frac{a(t)}{3}v^3 - \frac{b(t)}{2}\xi v^2] + \frac{e}{(a_2-r_1)\xi}v], \tag{56}$$

$$Q(v,t) = R(v,t)Q^{st}(v,t) = R(v,t)\exp[\frac{\alpha}{2(1-r_2)[(a_2-r_1)\xi]^2}[\frac{a'(t)}{12}v^4 - \frac{b'(t)}{6}\xi v^3]], \tag{57}$$

$$R(v,t) = S(v,t)R^{st}(v,t) = R(v,t)\exp[\frac{\alpha}{2(1-r_2)[(a_2-r_1)\xi]^3}[\frac{a''(t)}{60}v^5 - \frac{b''(t)}{24}\xi v^4]], \tag{58}$$

$$S(v,t) = R(v,t)S^{st}(v,t) = T(v,t)\exp[-\frac{\alpha}{2(1-r_2)[(a_2-r_1)\xi]^4}\frac{b'''(t)}{120}\xi v^5]. \tag{59}$$

By neglecting terms proportional to $1/\tau^3$ and taking the solutions as arbitrary functions of a variable $t+v/[(a_2-r_1)\xi]$, the arbitrary function $T(v,t)$ is set to $T(v,t) = \Theta[t+v/[(a_2-r_1)\xi]]$. Consequently, we find that

$$P(v,t) = T(v,t)S^{st}(v,t)R^{st}(v,t)Q^{st}(v,t)P^{st}(v,t) = \Theta[t+v/[(a_2-r_1)\xi]]S^{st}(v,t)R^{st}(v,t)Q^{st}(v,t)P^{st}(v,t). \tag{60}$$

We calculate the joint probability density from Eq. (55) and Eq. (60) as

$$P(\xi,v,t) = P(\xi,t)P(v,t)$$

$$= \exp[-\frac{5\alpha(a_2-r_1)t^4}{48(1-r_2)\tau}[1-\frac{12}{5}\frac{\tau}{t}]\xi^2 - \frac{\alpha t^3}{12(1-r_2)\tau}[1-6\frac{\tau}{t}]\xi v - \frac{3\alpha t^2}{8(a_2-r_1)(1-r_2)\tau}[1-\frac{2}{9}\frac{t}{\tau}]v^2]. \quad (61)$$

By using the inverse Fourier transform, the probability densities $P(v_r,t)$ and $P(v_\theta,t)$ are, respectively, presented by

$$P(v_r,t) = \frac{1}{2\pi}\int_{-\infty}^{+\infty} d\xi \exp(-i\xi v_r)P(\xi,0,t) = [\pi\frac{5\alpha(a_2-r_1)t^4}{12(1-r_2)\tau}[1-\frac{12}{5}\frac{\tau}{t}]]^{-1/2}\exp[-\frac{6(1-r_2)\tau v_r^2}{5\alpha(a_2-r_1)t^4}[1-\frac{12}{5}\frac{\tau}{t}]^{-1}], \quad (62)$$

$$P(v_\theta,t) = \frac{1}{2\pi}\int_{-\infty}^{+\infty} dv \exp(-ivv_\theta)P(0,v,t) = [2\pi\frac{3\alpha t^2}{4(a_2-r_1)(1-r_2)\tau}[1-\frac{2}{9}\frac{t}{\tau}]]^{-1/2}\exp[-\frac{2(a_2-r_1)(1-r_2)\tau v_\theta^2}{3\alpha t^2}[1-\frac{2}{9}\frac{t}{\tau}]^{-1}]. \quad (63)$$

The mean squared velocities for $P(v_r,t)$ and $P(v_\theta,t)$ from Eq. (62) and Eq. (63) are, respectively, given by

$$<v_r^2> = \frac{5\alpha(a_2-r_1)t^4}{24(1-r_2)\tau}[1-\frac{12}{5}\frac{\tau}{t}], \quad (64)$$

$$<v_\theta^2> = \frac{3\alpha t^2}{4(a_2-r_1)(1-r_2)\tau}[1-\frac{2}{9}\frac{t}{\tau}]. \quad (65)$$

### 3-2. $P(v_r,t)$ and $P(v_\theta,t)$ with the perturbative force in the long-time domain

In the long-time domain $t \gg \tau$, by applying the method (Eq. (30) - Eq. (34)) in the long-time domain for an active particle, we calculate the joint probability density as

$$P(\xi,v,t) = P(\xi,t)P(v,t)$$

$$= \exp[-\frac{\alpha(a_2-r_1)^2 t^3}{6(1-r_2)\tau}[1+\frac{3\tau}{2(a_2-r_1)t}]\xi^2 - \frac{\alpha(a_2-r_1)t^2}{2(1-r_2)}[1+\frac{\tau}{(a_2-r_1)t}]\xi v - \frac{\alpha t}{2(1-r_2)}v^2]. \quad (66)$$

For the inverse Fourier transform, the probability densities $P(v_r,t)$ and $P(v_\theta,t)$ from Eq. (66) are, respectively, presented as

$$P(v_r,t) = \frac{1}{2\pi}\int_{-\infty}^{+\infty} d\xi \exp(-i\xi v_r)P(\xi,0,t)$$

$$= [2\pi\frac{\alpha(a_2-r_1)^2 t^3}{3(1-r_2)\tau}[1+\frac{3\tau}{2(a_2-r_1)t}]]^{-1/2}\exp[-\frac{3(1-r_2)\tau v_r^2}{2\alpha(a_2-r_1)^2 t^3}[1+\frac{3\tau}{2(a_2-r_1)t}]^{-1}], \quad (67)$$

$$P(v_\theta,t) = \frac{1}{2\pi}\int_{-\infty}^{+\infty} dv \exp(-ivv_\theta)P(0,v,t) = [2\pi\frac{\alpha t}{(1-r_2)}]^{-1/2}\exp[-\frac{(1-r_2)v_\theta^2}{2\alpha t}]. \quad (68)$$

The mean squared deviations for $P(v_r,t)$ and $P(v_\theta,t)$ are, respectively, given by

$$<v_r^2> = \frac{\alpha(a_2-r_1)^2 t^3}{3(1-r_2)\tau}[1+\frac{3\tau}{2(a_2-r_1)t}], \quad (69)$$

$$<v_\theta^2> = \frac{\alpha t}{(1-r_2)}. \quad (70)$$

### 3-3. $P(v_r,t)$ and $P(v_\theta,t)$ with the perturbative force for $\tau = 0$

For $\tau = 0$, by applying the similar method (Eq. (41) – Eq. (42)), we also calculate the joint probability density as

$$P(\xi,v,t) = P(\xi,t)P(v,t) = \exp[-\frac{\alpha(a_2-r_1)^2 t^3}{6(1-r_2)\tau}[1+\frac{3\tau}{2(a_2-r_1)t}]\xi^2 - \frac{\alpha(a_2-r_1)t^2}{2(1-r_2)}\xi v - \frac{\alpha t}{2(1-r_2)}v^2]. \quad (71)$$

By using the inverse Fourier transform, $P(v_r,t)$ and $P(v_\theta,t)$ are, respectively, presented by

$$P(v_r,t) = \frac{1}{2\pi}\int_{-\infty}^{+\infty} d\xi \exp(-i\xi v_r)P(\xi,0,t) = [2\pi\frac{\alpha(a_2-r_1)^2 t^3}{3(1-r_2)\tau}[1+\frac{3\tau}{2(a_2-r_1)t}]]^{-1/2}\exp[-\frac{3(1-r_2)\tau v_r^2}{2\alpha(a_2-r_1)^2 t^3}[1+\frac{3\tau}{2(a_2-r_1)t}]^{-1}], \quad (72)$$

$$P(v_\theta,t) = \frac{1}{2\pi}\int_{-\infty}^{+\infty} dv \exp(-ivv_\theta)P(0,v,t) = [2\pi\frac{\alpha t}{(1-r_2)}]^{-1/2}\exp[-\frac{(1-r_2)v_\theta^2}{2\alpha t}]. \quad (73)$$

The mean squared velocities for $P(v_r,t)$ and $P(v_\theta,t)$ are, respectively, calculated as

$$<v_r^2> = \frac{\alpha(a_2-r_1)^2 t^3}{3(1-r_2)\tau}[1+\frac{3\tau}{2(a_2-r_1)t}], \quad (74)$$

$$<v_\theta^2> = \frac{\alpha t}{(1-r_2)}. \quad (75)$$

**3-4. Moment equation, kurtosis, correlation coefficient, and moment**

We derive the moment equation for $\mu_{m,n}$ of distribution $P(v_r,v_\theta,t)$ in Eq. (1) and Eq. (2) as follows:

$$\frac{d\mu_{m,n}}{dt} = -r_1 m\mu_{m,n} + a_2(n+1)\mu_{m,n} + \frac{a_1}{(1-r_2)}n\mu_{m+1,n-1} + \frac{\alpha b(t)}{(1-r_2)}mn\mu_{m-1,n-1} + \frac{\alpha a(t)}{(1-r_2)}(n-1)\mu_{m,n-2}. \quad (76)$$

Here,

$$\mu_{m,n} = \int_{-\infty}^{+\infty}dx\int_{-\infty}^{+\infty}dv x^m v^n P(v_r,v_\theta,t). \quad (77)$$

We also derive the moment equation for $\mu_{m,n}$ from the solution $P(v_r,v_\theta,t)$ with perturbative force in Eq. (4) and Eq. (5) as follows:

$$\frac{d\mu_{m,n}}{dt} = (r_1-a_2)m\mu_{m-1,n+1} + \frac{c}{(1-r_2)}n\mu_{m+3,n-1} - \frac{\alpha b(t)}{(1-r_2)}mn\mu_{m-1,n-1} + \frac{\alpha a(t)}{(1-r_2)}n(n-1)\mu_{m,n-2}. \quad (78)$$

The kurtoses for radial and tangential directions are, respectively, defined by

$$K_{v_r} = <v_r^4>/3<v_r^2>^2 - 1, \quad (79)$$

$$K_{v_\theta} = <v_\theta^4>/3<v_\theta^2>^2 - 1. \quad (80)$$

We lastly introduce the correlation coefficient as

$$\rho_{v_r,v_\theta} = <(v_r-\overline{v_r})(v_\theta-\overline{v_\theta})>/\sigma_{v_r}\sigma_{v_\theta}. \quad (81)$$

Here, we assume that the initial states for an active particle with the radial and tangential forces are $v_r = v_{r0}$, $v_\theta = v_{\theta 0}$, and $\overline{v_r}$, $\overline{v_\theta}$ denote velocities of joint probability density. The parameters $\sigma_{v_r}$ and $\sigma_{v_\theta}$ denote root-mean-squared velocities for the radial and tangential directions of joint probability density, respectively.

**Table 1.** Values of the kurtosis, the correlation coefficient, and the moment $\mu_{2,2}$ for an active particle with radial and tangential forces in the three-time domains.

| Time | $v_r, v_\theta$ | $K_{v_r}, K_{v_\theta}$ | $\rho_{v_r,v_\theta}$ | $\mu_{2,2}$ |
|---|---|---|---|---|
| $t \ll \tau$ | $v_r$ | $\frac{a_1^2 v_{r0}^4}{\alpha^2 r_1^4}t^{-6} + \frac{a_1 v_{r0}^2}{\alpha r_1^2}t^{-3}$ | $\frac{a_1 a_2(1-r_2)v_{r0}v_{\theta 0}}{\alpha^2 r_1^2}t^{-5}$ | $\frac{\alpha^2 r_1^2}{a_1(1-r_2)(r_1-a_2)\tau}t^4$ |
|  | $v_\theta$ | $\frac{a_2^2(1-r_2)^2\tau^2 v_{\theta 0}^4}{\alpha^2}t^{-4} + \frac{a_2(1-r_2)\tau v_{\theta 0}^2}{\alpha}t^{-2}$ |  |  |
| $t \gg \tau$ | $v_r$ | $\frac{a_1^2 v_{r0}^4}{\alpha^2 r_1^6(1-r_2)^2}t^{-4} + \frac{a_1 v_{r0}^2}{\alpha r_1^3(1-r_2)}t^{-2}$ | $\frac{a_1 v_{r0}v_{\theta 0}}{\alpha^2 r_1^3}t^{-3}$ | $\frac{\alpha^2 r_1^3}{a_1(r_1-a_2)\tau}t^3$ |
|  | $v_\theta$ | $\frac{(1-r_2)^2\tau^2 v_{\theta 0}^4}{\alpha^2}t^{-2} + \frac{(1-r_2)v_{\theta 0}^2}{\alpha}t^{-1}$ |  |  |
| $\tau = 0$ | $v_r$ | $\frac{a_1^4(1-r_2)^2 v_{r0}^4}{\alpha^2 r_1^6}t^{-4} + \frac{r_1(1-r_2)v_{r0}^2}{\alpha r_1^3}t^{-2}$ | $\frac{a_1^2(1-r_2)^2 v_{r0}v_{\theta 0}}{\alpha^2 r_1^3}t^{-3}$ | $\frac{\alpha^2 r_1^3}{a_1^2(1-r_2)^2(r_1-a_2)\tau}t^4$ |
|  | $v_\theta$ | $\frac{(1-r_2)^2\tau^2 v_{\theta 0}^4}{\alpha^2}t^{-2} + \frac{(1-r_2)v_{\theta 0}^2}{\alpha}t^{-1}$ |  |  |

**Table 2.** Values of the kurtosis, the correlation coefficient, and the moment $\mu_{2,2}$ for an active particle with perturbative force in three-time domains.

| Time | $v_r$, $v_\theta$ | $K_{v_r}$, $K_{v_\theta}$ | $\rho_{v_r,v_\theta}$ | $\mu_{2,2}$ |
|---|---|---|---|---|
| $t \ll \tau$ | $v_r$ | $\dfrac{r_1^2(1-r_2)^2\tau^2 v_{r0}^4}{\alpha^2(a_2-r_1)^2}t^{-8} + \dfrac{r_1(1-r_2)\tau v_{r0}^2}{\alpha(a_2-r_1)}t^{-4}$ | $\dfrac{r_1(1-r_2)^2\tau^2 v_{r0}v_{\theta 0}}{\alpha^2}t^{-6}$ | $\dfrac{\alpha^2(a_2-r_1)}{r_1(1-r_2)^2\tau}t^6$ |
|  | $v_\theta$ | $\dfrac{(a_2-r_1)^2(1-r_2)^2\tau^2 v_{\theta 0}^4}{\alpha^2}t^{-4} + \dfrac{(a_2-r_1)(1-r_2)\tau v_{\theta 0}^2}{\alpha}t^{-2}$ |  |  |
| $t \gg \tau$ | $v_r$ | $\dfrac{(1-r_2)^2\tau^2 v_{r0}^4}{\alpha^2(a_2-r_1)^4}t^{-6} + \dfrac{(1-r_2)\tau v_{r0}^2}{\alpha(a_2-r_1)^2}t^{-3}$ | $\dfrac{(1-r_2)^2\tau v_{r0}v_{\theta 0}}{\alpha^2(a_2-r_1)^2}t^{-4}$ | $\dfrac{\alpha^2(a_2-r_1)^2}{(1-r_2)^2\tau}t^5$ |
|  | $v_\theta$ | $\dfrac{(1-r_2)^2 v_{\theta 0}^4}{\alpha^2}t^{-2} + \dfrac{(1-r_2)v_{\theta 0}^2}{\alpha}t^{-1}$ |  |  |
| $\tau = 0$ | $v_r$ | $\dfrac{(1-r_2)^2\tau^2 v_{r0}^4}{\alpha^2(a_2-r_1)^4}t^{-6} + \dfrac{(1-r_2)\tau v_{r0}^2}{\alpha(a_2-r_1)^2}t^{-3}$ | $\dfrac{(1-r_2)^2\tau v_{r0}v_{\theta 0}}{\alpha^2(a_2-r_1)^2}t^{-4}$ | $\dfrac{\alpha^2(a_2-r_1)^2}{(1-r_2)^2\tau}t^5$ |
|  | $v_\theta$ | $\dfrac{(1-r_2)^2 v_{\theta 0}^4}{\alpha^2}t^{-2} + \dfrac{(1-r_2)v_{\theta 0}^2}{\alpha}t^{-1}$ |  |  |

## 4. Summary

In summary, we have derived a Fokker-Planck equation in directional active particle, subject to an exponentially correlated Gaussian force. We have obtained approximately the solution of the joint probability density by using double Fourier transforms. The mean squared velocity $<v_\theta^2(t)>$ ($<v_r^2(t)>$) for an active particle with both the radial and tangential forces and the perturbative force (with the radial and tangential forces) is proportional to $t^2$ ($t^2$) in the time domain $t \ll \tau$ (in $t \gg \tau$ and for $\tau = 0$). In particular, the mean squared tangential velocity $<v_\theta^2(t)>$ with both the radial and tangential forces and the perturbative force behaviors as the Gaussian diffusion in $t \gg \tau$ and for $\tau = 0$. This activity of an active particle leads to a super-diffusive distribution for the radial velocity $v_r(t)$ away from the Gaussian profile.

However, as has been studied so far, the mean squared displacement for an active self-propelled particle scales as ballistic transport $<x^2(t)> \sim t^2$ in $t \ll \tau$, and as $<x^2(t)> \sim t$ for $\tau = 0$ or $t \to \infty$ [37,38]. Compared with the self-propelled particle, our result of the mean squared tangential velocity $<v_\theta^2(t)>$ is matched with the same value to time $\sim t^2$, while the mean squared radial velocity $<v_r^2(t)>$ is the same as $<x^2(t)> \sim t$ in $t \gg \tau$ and for $\tau = 0$.

We obtain that the moment $\mu_{2,2}$ for joint probability density with radial and tangential forces is proportional to time $\sim t^3$ in $t \gg \tau$ and $\tau = 0$, and that for joint probability density with perturbative force is proportional to time $\sim t^4$ in $t \ll \tau$. These are consistent with our results, and the different values $\mu_{m,n}$ will be discussed elsewhere. In addition, the calculated values for the kurtosis, the correlation coefficient, and the moment from the moment equation are summarized in Table 1 and Table 2.

We will extend our model to the generalized Langevin equation or the force equation of motion with other forces [35-37]. The results will be compared and analyzed with theories, computer simulations, and experiments [38-46]. More detailed results for the probability density will be continuously published in other journals.

## Acknowledgments
This research was financially supported by the Ministry of Small and Medium-sized Enterprises (SMEs) and Startups (MSS), Korea, under the "Startup Growth Technology Development Project (R&D, 1425153351)" supervised by the Korea Technology and Information Promotion Agency for SMEs.